\documentclass[
    ,final            % use final for the camera ready runs
  ]
  {aipproc}

\layoutstyle{6x9}

\newcommand{\apj}{ApJ}
\newcommand{\apjs}{ApJS}

\begin{document}

\title{New astrometry and photometry for the companion candidates of CT Cha}


\author{Tobias O. B. Schmidt}{
  address={Astrophysikalisches Institut, Universit\"at Jena, Schillerg\"asschen 2-3, 07745 Jena, Germany, tobi@astro.uni-jena.de}
}

\author{Ralph Neuh\"auser}{
  address={Astrophysikalisches Institut, Universit\"at Jena, Schillerg\"asschen 2-3, 07745 Jena, Germany}
}

\author{Markus Mugrauer}{
  address={Astrophysikalisches Institut, Universit\"at Jena, Schillerg\"asschen 2-3, 07745 Jena, Germany}
}

\author{Ana Bedalov}{
  address={Faculty of Natural Sciences, University of Split, Teslina 12. 21000 Split, Croatia}
,altaddress={Astrophysikalisches Institut, Universit\"at Jena, Schillerg\"asschen 2-3, 07745 Jena, Germany} % additional 
}

\author{Nikolaus Vogt}{
  address={Departamento de F\'isica y Astronom\'ia, Universidad de Valpara\'iso, Avenida Gran Breta\~na 1111, Valpara\'iso, Chile}
  ,altaddress={Instituto de Astronom\'ia, Universidad Catolica del Norte, Avda.~Angamos 0610, Antofagasta, Chile} % additional visiting address
}

\begin{abstract}
In our ongoing search for close and faint companions around T Tauri stars in the Chamaeleon star-forming region, we recently (Schmidt et al.~2008b) presented direct observations and integral field spectroscopy of a new common proper motion companion to the young T-Tauri star and Chamaeleon member CT Cha and discussed its properties in comparison to other young, low-mass objects and to synthetic model spectra from different origins.
We now obtained for the first time direct H-Band observations of the companion CT Cha b and of another faint companion candidate (cc2) approximately 1.9 arcsec northeast of CT Cha using the Adaptive Optics (AO) instrument Naos-Conica (NACO) at the Very Large Telescope (VLT) of the European Southern Observatory (ESO) in February 2008.
From these data we can now exclude by 4.4 \& 4.8\,$\sigma$ that CT Cha b is a non-moving background object and find cc2 to be most likely a background star of spectral type $\leq$\,K4 with a proper motion of $\mu_{\alpha} \cos{\delta}$= -8.5\,$\pm$\,5.7\,mas/yr and $\mu_{\delta}$=\,12.0\,$\pm$\,5.6\,mas/yr, not consistent with being a member of the Cha I star-forming region.
\end{abstract}

\maketitle

%%%%%%%%%%%%%%%%%%%%%%%%%%%%%%%%%%%%%%%%%%%%
%% MAINMATTER
%%%%%%%%%%%%%%%%%%%%%%%%%%%%%%%%%%%%%%%%%%%%

\section{Introduction}

CT Cha (aka HM\,9), introduced in the 65th Name-List of Variable stars by 
Kholopov et al.~(1981), was originally found by Henize \& Mendoza~(1973) 
%as an emission-line star in Chamaeleon
%, exhibiting variations in its H$\alpha$\,line 
%from plate to plate and showing partial veiling \citep{1980AJ.....85..444R}.
and
%While the star was first classified as a T Tauri star by \citet{1987MNRAS.224..497W},
%it was 
later classified to be a classical T Tauri star by Weintraub (1990)
and Gauvin \& Strom (1992) from IRAS data (see Schmidt et al.~(2008b) and references therein for more information on CT Cha A).
%\citet{2000ApJ...534..838N} found 
%evidence of a silicate feature disk ($L_{sil}=10^{-2}\,L_{\odot}$, $L_{sil}/L_{\ast}=0.014$), 
%using ISO data.

%The variations in the H$\alpha$\,line were later interpreted as accretion signatures 
%when \citet{1998ApJ...495..385H} measured a mass accretion rate of log{\,\.{M}}$=-8.28\,M_{\odot}/a$.
%Additional variations in infrared \citep{1979MNRAS.187..305G} and optical photometry can 
%possibly be explained by surface features on CT Cha at a rotation period of 9.86 days,
%as found by \citet{1998A&AS..128..561B}.

%All 
%additional 
%properties of the K7 \citep{1988A&AS...76..347G} star CT Cha, such as
%its age ranging from 0.9 Myr \citep{2000ApJ...534..838N} to 3 Myr \citep{1993ApJ...416..623F},
%as well as its equivalent width of the lithium absorption line of W$_{\lambda}$(Li)\,=\,0.40\,$\pm$\,0.05\,\AA\,
%\citep{2007A&A...467.1147G}, its radial velocity of 15.1\,$\pm$\,0.1 km/s \citep{2006A&A...448..655J} 
%and proper motion (-21.3\,$\pm$\,4.6\,mas/yr and 6.3\,$\pm$\,4.5\,mas/yr in RA and Dec respectively \citep{2008arXiv0809.2812S}), are consistent with a very young member of the Cha I star-forming region, having an age of 2\,$\pm$\,2\,Myr.

We recently (Schmidt et al.~2008a \& 2008b), identified a very faint co-moving sub-stellar companion, just $\sim$~2.67 arcsec northwest of CT Cha corresponding to a projected separation of $\sim$ 440 AU as well as another companion candidate (cc2) $\sim$ 1.9 arcsec northeast of the primary star.
We could show that CT Cha A and CT Cha b form a common proper motion pair and that the companion is by $\geq$\,4\,$\sigma$ significance not a stationary background object. Our near-infrared spectroscopy yielded a temperature of 2600\,$\pm$\,250\,K for the companion and an optical extinction of $A_{\rm V}$=\,5.2\,$\pm$\,0.8\,mag, when compared to spectra calculated from Drift-Phoenix model atmospheres (Helling et al.~2008). We further demonstrated the validity of the model fits by comparison to several other well-known young sub-stellar objects. Due to a prominent Pa-$\beta$ emission in the J-band, accretion is probably still ongoing onto CT Cha b. From temperature and luminosity ($\log(L_{bol}/L_{\odot})$=\,--2.68\,$\pm$\,0.21), we derived a radius of R=\,2.20$_{-0.60}^{+0.81}$\,R$_{\mathrm{Jup}}$. We found a consistent mass of M=\,17\,$\pm$\,6\,M$_{\mathrm{Jup}}$ for CT Cha b from both its luminosity and temperature when placed on evolutionary tracks. Hence, CT Cha b is most likely a wide brown dwarf companion or possibly even a planetary mass object, depending on both the true mass of CT Cha b and the definition of planets.

\section{Astrometry}

\subsection{CT Cha b}

We observed CT Cha in three epochs in February 2006, March 2007 and in February 2008 (see Table \ref{table:1}).
All observations were done with the European Southern Observatory (ESO) Very Large Telescope (VLT) instrument Naos-Conica (NACO).

%%%%%%%%%%%%%%%%%%%%%%%%%%%%%%%%%%%%%%%%%%%%
%% SAMPLE TABLE
%%
%% Shows the use of \tablehead and \tablenote
%% macros
%%%%%%%%%%%%%%%%%%%%%%%%%%%%%%%%%%%%%%%%%%%%

\begin{table}
\begin{tabular}{ccccccccc}
\hline
JD - 2453700 $[\mathrm{days}]$ & Date of observation & DIT [s] & NDIT & No.~of images & Filter\\
\hline
\ \ 83.54702 & 17 Feb 2006 & 1.5 & 25 & 20 & Ks \\
   460.62535 &  1 Mar 2007 & 4   & 15 & 21 & J  \\
   461.63001 &  2 Mar 2007 & 4   & 7  & 30 & Ks \\
   815.81036 & 19 Feb 2008 & 1   & 60 &  5 & H  \\
\hline
\end{tabular}
\caption{VLT/NACO observation log. Remark: Each image consists of the number of exposures given in column 4 times the individual integration time given in column 3.
}
\label{table:1}
\end{table}

\begin{figure}
  \includegraphics[height=.275\textheight]{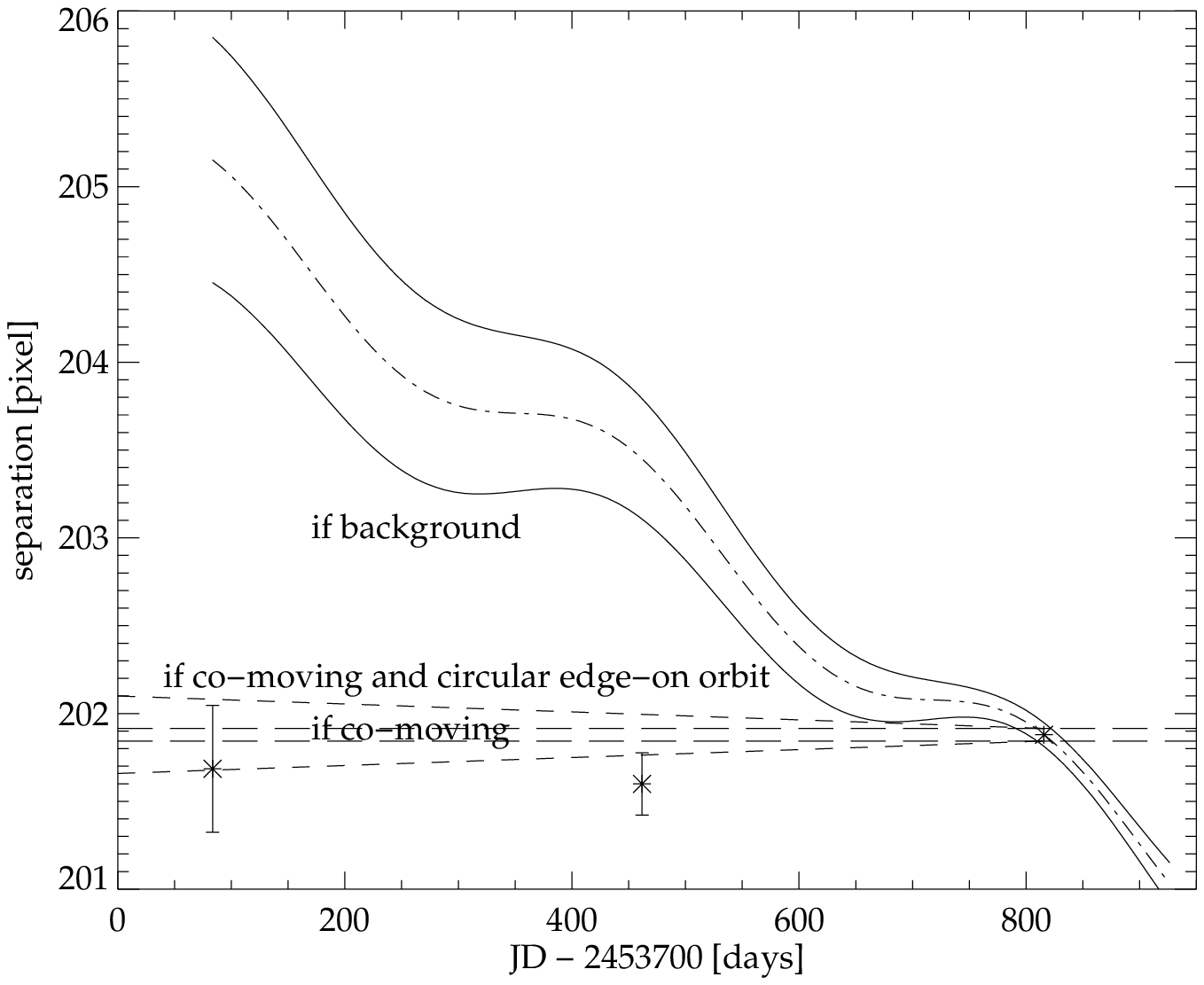}
  \includegraphics[height=.275\textheight]{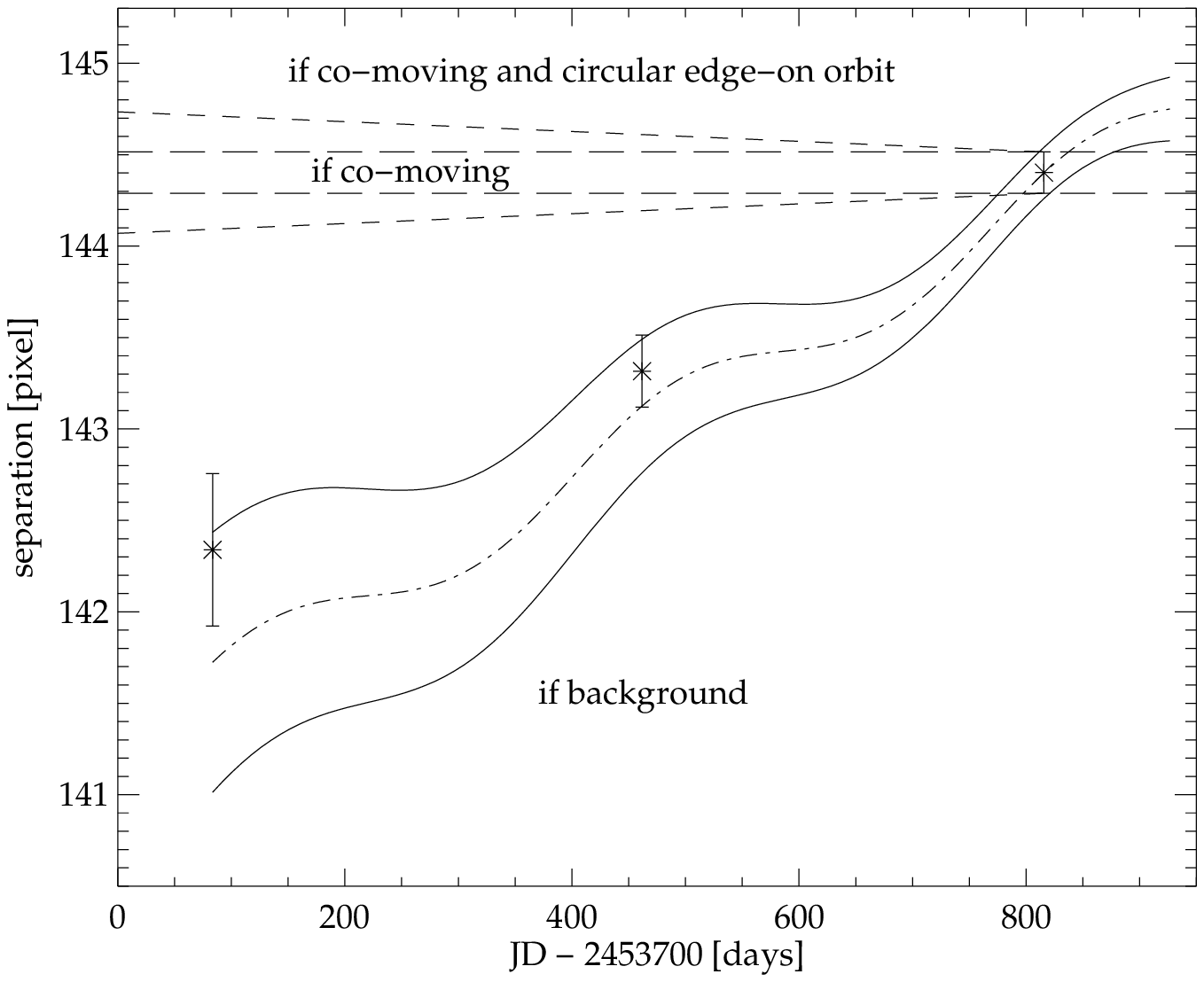}
  \caption{Observed separation between CT Cha A and CT Cha b (\textit{left}) and cc2 (\textit{right}). Our measurements from 2006, 2007, 2008 are shown. The long dashed lines enclose the area for constant separation, as expected for a co-moving object. The dash-dotted line is the change expected if CT Cha b / cc2 is a non-moving background star. The opening cone enclosed by the continuous lines its estimated errors. The waves of this cone show the differential parallactic motion that has to be taken into account if the other component is a non-moving background star with negligible parallax. The opening short-dashed cone is for the combination of co-motion and the maximum possible orbital motion for a circular edge-on orbit.}
\label{fig:1}
\end{figure}

To check again for the common proper motion of CT Cha b we have repeated the analysis done in Schmidt et al.~(2008b) and included the new H-Band epoch listed as last entry in  Table \ref{table:1}.
From Fig.~\ref{fig:1} (left), we can now exclude by 4.4 \& 4.8\,$\sigma$ that CT Cha b is a non-moving background object.

%%%%%%%%%%%%%%%%%%%%%%%%%%%%%%%%%%%%%%%%%%%%
%% Sample figure:
%%
%% The option [height=...] scales the picture to the given height,
%% without it it would be printed at its nominal size
%%%%%%%%%%%%%%%%%%%%%%%%%%%%%%%%%%%%%%%%%%%%

\subsection{Companion Candidate 2 (cc2)}

\begin{table}
\centering
\begin{tabular}{lccr@{\,$\pm$\,}lccc}
\hline
JD - 2453700 & Pixel scale & Separation  & \multicolumn{2}{c}{PA$^a$ } & Change in sepa- & Change in \\
$[\mathrm{days}]$ & [mas/Pixel] & [arcsec] &  \multicolumn{2}{c}{[$\deg$]} & ration [pixel] & PA$^a$ [$\deg$]\\
\hline
 \ \ 83.54702 & 13.24 $\pm$ 0.18 & 2.670 $\pm$ 0.036 & 300.71 & 1.24 & -0.19 $\pm$ 0.36 & 0.219 $\pm$ 0.166 \\
   461.63001  & 13.24 $\pm$ 0.19 & 2.670 $\pm$ 0.038 & 300.68 & 1.32 & -0.28 $\pm$ 0.18 & 0.183 $\pm$ 0.082 \\
   815.81036  & 13.24 $\pm$ 0.20 & 2.674 $\pm$ 0.040 & 300.50 & 1.40 & $^b$ & $^b$ \\
\hline
\end{tabular}
\caption{Absolute and relative astrometric results for CT Cha A and CT Cha b. (a) PA is measured from N over E to S. (b) Relative changes are given relative to epoch 3 at JD 2454515.81036.}
\label{table:2}
\end{table}

To check for the common proper motion of the tentative second companion (cc2) of CT Cha, we also repeated the analysis done in Schmidt et al.~(2008b) for CT Cha b for this object and included the new H-Band epoch (see Table \ref{table:1}).
As can be seen in Fig.~\ref{fig:1}, the companion candidate cc2 is moving like a stationary background object in separation from CT Cha A and, hence, is not a low-mass companion. Given the change in separation and position angle, the proper motion of CT Cha cc2 $\mu_{\alpha} \cos{\delta}$= -8.5\,$\pm$\,5.7\,mas/yr and $\mu_{\delta}$=\,12.0\,$\pm$\,5.6\,mas/yr is by $\sim$\,3\,$\sigma$
not consistent with the median proper motion of 61 members of the Cha I star-forming region of $\mu_{\alpha} \cos{\delta}$= \mbox{-21}\,$\pm$\,1\,mas/yr and $\mu_{\delta}$=\,2.0\,$\pm$\,1\,mas/yr (Luhman et al. 2008).

\begin{table}
\centering
\begin{tabular}{cr@{\,$\pm$\,}lr@{\,$\pm$\,}lr@{\,$\pm$\,}lr@{\,$\pm$\,}lr@{\,$\pm$\,}lr@{\,$\pm$\,}l}
\hline
      &\multicolumn{2}{c}{}&\multicolumn{2}{c}{CT Cha b}&\multicolumn{2}{c}{}&\multicolumn{2}{c}{}&\multicolumn{2}{c}{CT Cha cc2}&\multicolumn{2}{c}{}\\
Epoch & \multicolumn{2}{c}{J-band} & \multicolumn{2}{c}{H-band} & \multicolumn{2}{c}{Ks-band}& \multicolumn{2}{c}{J-band} & \multicolumn{2}{c}{H-band} & \multicolumn{2}{c}{Ks-band} \\
\hline
2006   & \multicolumn{2}{c}{}      & \multicolumn{2}{c}{}    & 14.95  & 0.30 & \multicolumn{2}{c}{}      & \multicolumn{2}{c}{}    & 17.36  & 0.31 \\
2007   & 16.61  & 0.30    & \multicolumn{2}{c}{}    & 14.89  & 0.30 & 18.15  & 0.32    &  \multicolumn{2}{c}{}  & 17.27  & 0.31 \\
2008   & \multicolumn{2}{c}{}      & 15.56  & 0.30  & \multicolumn{2}{c}{}   & \multicolumn{2}{c}{}      & 17.48  & 0.32  & \multicolumn{2}{c}{}   \\
\hline
\end{tabular}
\caption{Apparent magnitudes of CT Cha b and CT Cha cc2. See Table \ref{table:1} for JDs of the observations.}
\label{table:3}
\end{table}

\section{Photometry}

\begin{figure}
  \includegraphics[height=.33\textheight]{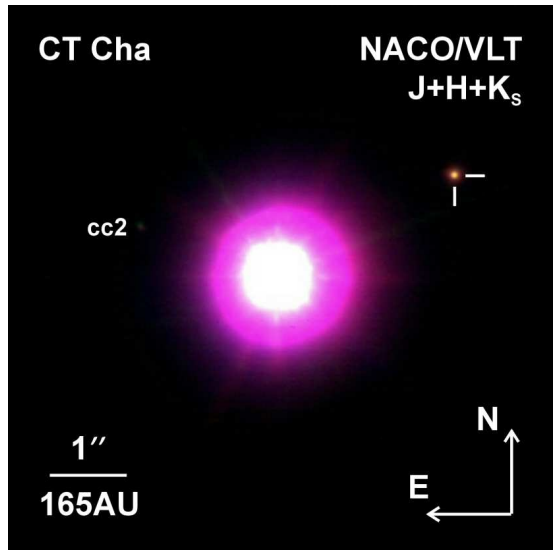}
  \caption{Artificial color-composite image from our J-, H- and Ks-band images taken in 2007 and 2008. CT Cha b was found $\sim$\,2.7 arcsec northwest (marked) of the classical T Tauri star CT Cha A. Northeast of CT Cha A the background star cc2 has currently a separation of $\sim$\,1.9 arcsec.}
\label{fig:2}
\end{figure}

In addition we compute from our NACO measurements also H-band photometry from the flux ratio of the companion candidates relative to CT Cha A (see Table \ref{table:3}). From the J-, H- and Ks-band magnitudes we can now present an artificial color-composite image, shown in Fig.~\ref{fig:2}. If we assume the same amount of extinction for cc2 as we found for CT Cha A, i.e. $A_{V}\sim$\,1.3\,mag (Schmidt et al.~2008b) or possibly more, as the star might be a background star, we find a spectral type of $\leq$\,K4 from the J-, H-, and K-band photometry given in Table \ref{table:3} in comparison to the colors of main-sequence stars given in Kenyon \& Hartmann (1995), hence, at the main-sequence, a distance of $\geq$\,350\,pc.

%%%%%%%%%%%%%%%%%%%%%%%%%%%%%%%%%%%%%%%%%%%%%%%%
%% BACKMATTER
%%%%%%%%%%%%%%%%%%%%%%%%%%%%%%%%%%%%%%%%%%%%%%%%

\begin{theacknowledgments}
TOBS would like to thank Evangelisches Studienwerk e.V. Villigst for financial support. NV acknowledges support by FONDECYT (grant 1061199) and Universidad de Valparaiso (grant 07/2007). AB and RN wish to acknowledge support from the German National Science Foundation (Deutsche Forschungsgemeinschaft, DFG), grants NE 515/13-1 and 13-2.
\end{theacknowledgments}



\bibliographystyle{aipprocl} % if natbib is missing

%\bibliography{references.bib}

\IfFileExists{\jobname.bbl}{}
 {\typeout{}
  \typeout{******************************************}
  \typeout{** Please run "bibtex \jobname" to optain}
  \typeout{** the bibliography and then re-run LaTeX}
  \typeout{** twice to fix the references!}
  \typeout{******************************************}
  \typeout{}
 }

\end{document}